\documentclass[a4paper]{jpconf}
\usepackage{amsmath}
\usepackage{graphicx}
\usepackage{hyperref}
\begin{document}
%\title{Magnetic and orbital instabilities of an extension of the SU(4) Kondo effect to the two-dimensional lattice}
%\title{\textcolor{red}{Magnetic and orbital instabilities in the two-dimensional SU(4) Kondo lattice}}
\title{Magnetic and orbital instabilities in a lattice of SU(4) organometallic Kondo complexes}

\author{A. M. Lobos$^1$ and A. A. Aligia$^2$ }
\address{
$^1$Joint Quantum Institute and Condensed Matter Theory Center, Department of
Physics, University of Maryland, College Park, Maryland 20742, USA \\
$^2$Centro At\'omico Bariloche and Instituto Balseiro, Comisi\'on Nacional de
Energ\'{\i}a At\'omica, 8400 Bariloche, Argentina}

\ead{aligia@cab.cnea.gov.ar}

\begin{abstract}
Motivated by experiments of scanning tunneling spectroscopy (STS) on self-assembled networks of iron(II)-phtalocyanine (FePc) molecules deposited on a clean Au(111) surface [FePc/Au(111)] and its explanation in terms of the extension of the impurity SU(4) Anderson model to the lattice in the Kondo regime, we study the competition between the Kondo effect and the magneto-orbital interactions occurring in FePc/Au(111). We explore the quantum phases and critical points of the model using a large-$N$ slave-boson method in the mean-field approximation. The
SU(4) symmetry in the impurity appears as a combination of the usual spin and an orbital pseudospin arising from the degenerate $3d_{xz}$ and $3d_{yz}$ orbitals in the Fe atom. In the case of the lattice, our results show that the additional orbital degrees of freedom crucially modify the low-temperature phase diagram, and induce new types of orbital interactions among the Fe atoms, which can potentially stabilize exotic quantum phases with magnetic and orbital order.
The dominant instability corresponds to spin ferromagnetic and orbital antiferromagnetic order.\end{abstract}

\section{Introduction}

Organometallic complexes containing magnetic centers are currently under intense investigation for their 
potential uses as building blocks for nanotechnologies. Low-dimensional magnetic nanostructures have a 
high potential for applications in spintronics, magnetic recording and sensing devices \cite{boga,bader}  
These systems also offer a unique platform to study exotic phases of matter. In particular,
Minamitani {\it et al.} \cite{mina} have shown that the Kondo effect observed in isolated iron(II)
phtalocyanine (FePc) molecules deposited on top of clean Au(111) [FePc/Au(111)]
(in the most usual on-top configuration) is a new realization of the SU(4) Kondo model,
in which not only the spin degeneracy but also the orbital degeneracy 
between $3d_{xz}$ and $3d_{yz}$ orbitals of Fe play a role ($z$ is the direction normal to the 
surface). The SU(4) Kondo effect manifests itself  as a dip at the Fermi energy in the differential
conductance observed in scanning tunneling spectroscopy (STS). The half width at half maximum
of the dip is about $T_K \sim$ 0.4 meV, with $T_K$ the Kondo temperature.

In another set of experiments, a self-organized square lattice of FePc/Au(111)  
as well as small clusters were studied by STS \cite{tsuka}.
It was found that as for the isolated molecule, a single dip in the differential
conductance remains for the molecules that lie at the corners of the clusters.  However, increasing the coordination, the peak tends to split and for the lattice,
a clear splitting of approximately 2 meV becomes apparent. Romero and  the present authors were able to explain
these experimental results, using a natural generalization of the single-impurity SU(4)
Anderson model to the lattice, including hoppings between effective orbitals of 
nearest-neighbor molecules respecting the symmetry of these orbitals \cite{mole}.
The relevant effective orbitals (i.e., the closest to the Fermi level), are essentially the $3d_{xz}$ and $3d_{yz}$ of Fe, 
with some admixture of other orbitals of the molecule. The splitting of the Kondo dip is a 
consequence of the anisotropy in the hopping for a given effective orbital.

Preliminary results suggests that the square lattice of FePc/Au(111)FePc 
is not far from a magnetic instability \cite{mole}.
Ferromagnetic order was observed in a two-dimensional layer of organic molecules absorbed on
graphene \cite{gar} and in metal-organic networks on Au surfaces \cite{umba,abdu}.
In addition, magnetic and orbital ordering are intertwined \cite{kk,gusm}.

In this work, starting from the orbitally degenerate Hubbard-Anderson model that describes 
the observed STS in different arrays of FePc molecules \cite{mole}, 
we perform a Schrieffer-Wolff transformation
to obtain an effective model which includes spin and orbital interactions.
Solving this model in a slave-boson mean-field approximation (SBMFA), we obtain the critical values of the interactions
leading to different symmetry-breaking magnetic and orbital instabilities.   The dominant one
turns out to be a spin ferromagnetic and orbital antiferromagnetic order.
We show that the effect of the Ruderman-Kittel-Kasuya-Yosida (RKKY) interactions is small and can be neglected.

\section{The Hubbard-Anderson model}

Our starting model was derived Ref. \cite{mole} and discussed in detail in
the supplemental material of that paper. The Hamiltonian $H=H_{1}+H_{2}$ can
be separated into one-body ($H_{1}$) and two-body ($H_{2}$) parts. The first
one can be written as

\begin{eqnarray}
H_{1}&=&\sum_{ij}^{N} [ E_{h}n_{\mathbf{r}_{ij}}-\sum_{\sigma \nu }\left(
t_{2}h_{\mathbf{r}_{ij},\sigma }^{\nu \dagger }h_{\mathbf{r}_{ij}\pm 
\mathbf{a}_{\nu },\sigma }^{\nu }+t_{1}h_{\mathbf{r}_{ij},\sigma }^{\bar{\nu}\dagger
}h_{\mathbf{r}_{ij}\pm \mathbf{a}_{\nu },\sigma }^{\bar{\nu}}\right)
+\sum_{\xi \sigma \nu }\epsilon _{\xi }c_{\mathbf{r}_{ij},\xi ,\sigma }^{\nu
\dagger }c_{\mathbf{r}_{ij},\xi ,\sigma }^{\nu } \nonumber \\
&+&V\sum_{\xi \sigma \nu
}\left( h_{\mathbf{r}_{ij},\sigma }^{\nu \dagger }c_{\mathbf{r}_{ij},\xi
,\sigma }^{\nu }+\rm{H.c.}\right) ] .  \label{h1}
\end{eqnarray}
Here the operators $h_{\mathbf{r}_{ij},\sigma }^{\nu }$ annihilate a hole
(create an electron) in the state $\vert \nu _{\mathbf{r}_{ij},\sigma
}\rangle$, where $\nu =\left( x,y\right) $ denotes one of the two
orbitally degenerate molecular states
%(defined conveniently so that $\langle x_{% \mathbf{r}_{ij},\sigma }\right\vert H\left\vert y_{\mathbf{r}_{lm},\sigma}\right\rangle =0$, for all $\mathbf{r}_{ij},\mathbf{r}_{lm}$) 
with spin $\sigma $ at site with position $\mathbf{r}_{ij}$ of the square lattice. 
$n_{\mathbf{r}_{ij}}=\sum_{\sigma \nu }n_{\mathbf{r}_{ij},\sigma }^{\nu }$, with 
$n_{\mathbf{r}_{ij},\sigma }^{\nu }=h_{\mathbf{r}_{ij},\sigma }^{\nu \dagger
}h_{\mathbf{r}_{ij},\sigma }^{\nu }$ is the total number of holes at the
molecule lying at site $ij$. The lattice vectors $\mathbf{a}_{\nu }$ connect
nearest-neighbor sites, $\bar{x}=y$, and$\ \bar{y}=x$. The operator $c_{\mathbf{r}_{ij},\xi ,\sigma }$ 
annihilates a conduction
hole with spin $\sigma $ and quantum number $\xi $ at position $\mathbf{r}_{ij}$. 
The first and second terms of \ $H_{1}$ describe the molecular
states and the hopping between them. The hopping $t_{2}$ between $x$ ($y$)
orbitals in the $x$ ($y$) direction is larger than the hopping $t_{1}$
between $x$ ($y$) orbitals in the $y$ ($x$) direction. The third term of \ $%
H_{1}$ corresponds to a band of bulk and surface conduction electrons of the
substrate and the last term is the hybridization between molecular and
conduction states.

The interaction term can be written as \cite{gusm,kroll}

\begin{eqnarray}
H_{2}&=&\sum_{ij}^{N} \left[ U\sum_{\nu }n_{\mathbf{r}_{ij},\uparrow }^{\nu }n_{%
\mathbf{r}_{ij},\downarrow }^{\nu }+\sum_{\sigma \sigma^\prime }[(U-2J_{H})n_{\mathbf{r%
}_{ij},\sigma }^{x}n_{\mathbf{r}_{ij},\sigma ^{\prime }}^y+J_{H}h_{\mathbf{r}_{ij},\sigma }^{x \dagger} h_{\mathbf{r}_{ij},\sigma
^{\prime }}^{y \dagger} h_{\mathbf{r}_{ij},\sigma ^{\prime }}^x h_{\mathbf{r}_{ij},\sigma }^y] \right. \nonumber  \\
&+&\left. J_{H}\sum_{\nu }h_{\mathbf{r}%
_{ij},\uparrow }^{\nu \dagger }h_{\mathbf{r}_{ij},\downarrow }^{\nu \dagger
}h_{\mathbf{r}_{ij},\downarrow }^{\bar{\nu}}h_{\mathbf{r}_{ij},\uparrow }^{%
\bar{\nu}} \right] .  \label{h2}
\end{eqnarray}
For the system we are considering, the occupation of the molecular states is
nearly one-hole per site ($\left\langle n_{\mathbf{r}_{ij}}\right\rangle
\sim 1$) and the main effect of $H_{2}$ is to inhibit double occupancy.
Therefore as a first approximation one can take $U\longrightarrow +\infty $
and therefore neglect the Hund rules exchange $J_{H}$ in comparison with $U$. 
In this case, the interaction takes the simpler form
$H_{2}^{s}=\frac{U}{2}\sum_{ij}^{N}n_{\mathbf{r}_{ij}}\left( n_{\mathbf{r}_{ij}}-1\right)$.

\section{Symmetry of $H_{i}$}

For the case of one molecule only, the resulting impurity Anderson
Hamiltonian has SU(4) symmetry, which in simple terms means that
permutations of the four states $\left\vert \nu _{\sigma }\right\rangle $
leave the Hamiltonian invariant. The fifteen generators of the SU(4) symmetry are
three trivial diagonal matrices, six permutations of two states and other
six permutations with a change of phases for the permuted states \cite{sba}.
The twelve non trivial generators can also be written as a generalization of
the raising and lowering operators for SU(2) \cite{li}. Specifically, for
the impurity Anderson model they are $S_{\nu \sigma }^{\nu ^{\prime }\sigma ^{\prime
}}=\left( h_{\sigma }^{\nu \dagger }h_{\sigma ^{\prime }}^{\nu ^{\prime }} + \sum_\xi c_{\xi,\sigma }^{\nu \dagger }c_{\xi,\sigma ^{\prime }}^{\nu ^{\prime }}\right)$ for $%
\nu \sigma \neq \nu ^{\prime }\sigma ^{\prime }$ (note that the  conduction electron degrees of freedom must be taken into account to keep the SU(4) symmetry of the total system.)

For the lattice, the simplest generalization of these generators leads to 
$S_{\nu \sigma }^{\nu ^{\prime }\sigma ^{\prime }}=\sum_{ij}^{N}\left( h_{\mathbf{r}%
_{ij},\sigma }^{\nu \dagger }h_{\mathbf{r}_{ij},\sigma ^{\prime }}^{\nu
^{\prime }}+ \sum_\xi c_{\mathbf{r}%
_{ij},\xi,\sigma }^{\nu \dagger }c_{\mathbf{r}%
_{ij},\xi,\sigma ^{\prime }}^{\nu ^{\prime }}\right)$. All these generators commute with $H_{2}^{s}$, but those with $%
\nu \neq \nu ^{\prime }$ commute with $H_{1}$ only in the particular case 
$t_{2}=t_{1}$, which seems incompatible with the observed STS in the square
lattice of FePc/Au(111) \cite{mole}. In the general case, $t_{2}\neq t_{1}$%
, $H_{1}$ has however SU(4) symmetry with the following non-trivial generators $S_{\nu
\sigma }^{\nu \bar{\sigma}}=\sum_{ij}^{N}\left(h_{\mathbf{r}_{ij},\sigma }^{\nu
\dagger }h_{\mathbf{r}_{ij},\bar{\sigma}}^{\nu }+ \sum_\xi c_{\mathbf{r}%
_{ij},\xi,\sigma }^{\nu \dagger }c_{\mathbf{r}%
_{ij},\xi,\bar{\sigma}}^{\nu}\right)$ and 
$S_{\nu \sigma }^{\bar{\nu}\sigma ^{\prime }}=\sum_{ij}^{N}\left(h_{\mathbf{r}_{ij},\sigma }^{\nu
\dagger }h_{R\mathbf{r}_{ij},\sigma ^{\prime }}^{\bar{\nu}}+ \sum_\xi c_{\mathbf{r}%
_{ij},\xi,\sigma }^{\nu \dagger }c_{R\mathbf{r}%
_{ij},\xi,\sigma ^{\prime }}^{\bar{\nu}}\right)$,
where $R$ is the reflection that permutes $x$ and $y$ (it is an element of
the point group $C_{4v}$ of the system). It can be verified easily that
these generators commute with $H_{1}$. However, inclusion of $H_{2}$ reduces
the symmetry to spin SU(2) times orbital Z$_2\times$U(1). Only the term of $H_{2}^{s}$
with $\mathbf{r}_{ij}=0$ commutes with the $S_{\nu \sigma }^{\bar{\nu}\sigma
^{\prime }}$ generators which contain $R$. Nevertheless, in a Fermi liquid,
the interaction becomes irrelevant at the Fermi energy and we expect that
SU(4) is an emergent symmetry at low energies \cite{bat} if there is not a
symmetry breaking (a magnetic or orbital instability). In fact, in  the SBMFA,
%slave-boson mean-field approximation 
where the action is reduced to an
effective non-interacting one  near the Fermi energy \cite{mole}
or in a dynamical-mean field approximation in which the interaction is
treated exactly at one site in an effective medium, the effective model has
SU(4) symmetry if the symmetric form $H_{2}^{s}$ is taken.

\section{Effective generalized Heisenberg interactions}

\vspace{0.5cm}When two nearest-neighbor sites are singly occupied and if 
$U\gg t_{i}$ (as it seems to be the case for FePc/Au(111) \cite{mole}, 
the hopping terms connecting these sites can be eliminated by means
of a canonical transformation, in a similar fashion as the $t-J$ model is
derived from the Hubbard model. This leads to an effective exchange model for spins and orbitals, as in the Kugel-Khomskii model \cite%
{kk,gusm}. For simplicity we write first the result using the SU(4)
symmetric form of the interaction $H_{2}^{s}$. 
%ACA The main change that takes place if the more realistic form $H_{2}$ [Eq. (\ref{h2})]
%is used is that $U$ in the denominators should be replaced by $U-3J_{H}$
%when the spins of the nearest-neighbor sites form a triplet, favoring ferromagnetic order. 
After a lengthy but straightforward algebra we obtain

\begin{eqnarray}
H_{H}& =\sum_{i}\sum_{\nu }\left[ \frac{4t_{2}^{2}}{U}\mathbf{S}_{i}^{\nu
}\cdot \mathbf{S}_{i+\nu }^{\nu }+\frac{4t_{1}^{2}}{U}\mathbf{S}_{i}^{\bar{%
\nu}}.\mathbf{S}_{i+\nu }^{\bar{\nu}}\right] +\sum_{i}\sum_{\nu }\left[ 
\frac{t_{2}^{2}+t_{1}^{2}}{4U}\left( 4T_{i}^{z}T_{i+\nu }^{z}-3\right) %
\right]   \nonumber \\
& +\sum_{i}\sum_{\nu }\frac{2t_{1}t_{2}}{U}\left( T_{i}^{-}T_{i+\nu
}^{+}+T_{i}^{+}T_{i+\nu }^{-}\right) \left( \frac{1}{2}+2\mathbf{S}_{i}\cdot 
\mathbf{S}_{i+\nu }\right) ,  \label{hh}
\end{eqnarray}%
where $\mathbf{S}_{i}^{\nu }=\sum_{\alpha \beta }h_{i\alpha }^{\nu \dagger }{%
\mathbf{\sigma }}_{\alpha \beta }h_{i\beta }^{\nu }/2$ is the spin of
the orbital $\nu $ at site $i$, and $\mathbf{S}_{i}=\sum_{\nu }\mathbf{S}%
_{i}^{\nu }$. The operator $\mathbf{T}_{i}$ denotes the orbital SU(2) pseudospin (with the identification of $x$ for pseudospin up). The subscript $i+\nu $ denotes the nearest neighbor of $i$
in the $+ \nu $ direction. Note that in the case $t_{1}=t_{2}=t$, this Hamiltonian reduces to

\begin{equation}
H_{H}^{SU(4)}=\frac{2t^{2}}{U}\sum_{i}\sum_{\nu }\left( 2\mathbf{S}_{i}\cdot 
\mathbf{S}_{i+\nu }+\frac{1}{2}\right) \left( 2\mathbf{T}_{i}\cdot \mathbf{T}%
_{i+\nu }+\frac{1}{2}\right) +\rm{cte}.  \label{hhsu4}
\end{equation}%
where we have used that $\mathbf{S}_{i}^{\nu }\cdot \mathbf{S}_{i+\nu }^{\nu
}+\mathbf{S}_{i}^{\bar{\nu}}\cdot \mathbf{S}_{i+\nu }^{\bar{\nu}}=\mathbf{S}%
_{i}\cdot \mathbf{S}_{i+\nu }\left( 2T_{i}^{z}T_{i+\nu }^{z}+1/2\right) $.
This Hamiltonian is a sum of products of a spin SU(2) invariant form (first
factor) times a pseudospin SU(2) invariant (last factor). Thus, it is
explicitly SU(2)$\times $SU(2) invariant. However, it has been shown \cite%
{li} that the symmetry of $H_{H}^{SU(4)}$ is actually SU(4), which is larger
than SU(2)$\times $SU(2).

When $t_{1}$ and $t_{2}$ are very different, as in the realistic case for
for FePc/Au(111) \cite{mole}, the first two terms of $H_{H}$ Eq. (\ref{hh})
are the most important ones. The first one is optimized for orbital
ferromagnetic and spin antiferromagnetic 
order, while the second one favors orbital antiferromagnetic order. In a classical picture, the energy of both orders would be the same, $%
-(t_{1}^{2}+t_{2}^{2})/U$ per site. However when Hund rules are included
[the form $H_{2}$, Eq. (\ref{h2}) is used for the interaction] the spin
ferromagnetic order is favored. Projecting over intermediate double occupied
triplet states, the dominant term of $H_{H}$ takes the form

\begin{equation}
H_{H}^{d}=\frac{J}{2}\sum_{\mathbf{r}_{ij},\mathbf{a}}\left( -1
+4T_{\mathbf{r}_{ij}}^{z}T_{\mathbf{r}_{ij}+\mathbf{a}}^{z}\right) \left( \frac{3}{4}+
\mathbf{S}_{\mathbf{r}_{ij}}.\mathbf{S}_{\mathbf{r}_{ij}+\mathbf{a}}\right), 
\hspace{1cm}{\rm }J=\frac{t_{1}^{2}+t_{2}^{2}}{2(U-3J_{H})}  \label{hd}
\end{equation}

\section{Instabilities due to $H_{H}$}

The simplest SBMFA of  $H$ \cite{mole} is not
enough to treat the magnetic and orbital instabilities. As it is usually
done in mean-field treatments of the Kondo lattice, in which the RKKY
interaction should be included explicitly \cite{coq}, in this section we
consider $H+H_{H}^{d}$ within the SBMFA to study the instabilities induced
by $H_{H}^{d}$. In this approximation, the limit $U\rightarrow +\infty $ is
taken with a constraint of forbidden double occupancy. However, the effects
of a finite $U$ are considered explicitly in $H_{H}$.

As before \cite{mole}, the hole operators are written in terms of auxiliary particles as $%
h_{\mathbf{r}_{ij},\sigma }^{\nu }=b_{\mathbf{r}_{ij}}^{\dagger }f_{\mathbf{%
r}_{ij},\sigma }^{\nu }$. The spin and pseudospin operators take the form 

\begin{equation}
\mathbf{S}_{\mathbf{r}_{ij}}\equiv -\frac{1}{2}\sum_{\sigma ,\nu }f_{\mathbf{%
r}_{ij},\sigma }^{\nu \dagger }\hat{\mathbf{\sigma }}_{\sigma ,\sigma
^{\prime }}f_{\mathbf{r}_{ij},\sigma ^{\prime }}^{\nu } ,\hspace{0.5cm} \mathbf{T}_{%
\mathbf{r}_{ij}}\equiv -\frac{1}{2}\sum_{\sigma ,\nu }f_{\mathbf{r}%
_{ij},\sigma }^{\nu \dagger }\hat{\tau}_{\nu ,\nu ^{\prime }}f_{\mathbf{r}%
_{ij},\sigma }^{\nu ^{\prime }},  \label{st}
\end{equation}%
We also define the mixed operator 
%\begin{equation}U_{\mathbf{r}_{ij}}^{z}\equiv S_{\mathbf{r}_{ij}}^{z}T_{\mathbf{r}_{ij}}^{z}=\frac{1}{4}\left( f_{\mathbf{r}_{ij},\uparrow }^{x\dagger }f_{\mathbf{r}_{ij},\uparrow }^{x}-f_{\mathbf{r}_{ij},\uparrow }^{y\dagger }f_{\mathbf{r}_{ij},\uparrow }^{y}-f_{\mathbf{r}_{ij},\downarrow }^{x\dagger }f_{\mathbf{r}_{ij},\downarrow }^{x}+f_{\mathbf{r}_{ij},\downarrow }^{y\dagger }f_{\mathbf{r}_{ij},\downarrow }^{y}\right) =\frac{1}{4}\sum_{\sigma ,\nu }\eta \left(\sigma \right) \eta \left( \nu \right) f_{\mathbf{r}_{ij},\sigma }^{\nu\dagger }f_{\mathbf{r}_{ij},\sigma }^{\nu }\end{equation}
$U_{\mathbf{r}_{ij}}^{z}\equiv S_{\mathbf{r}_{ij}}^{z}T_{\mathbf{r}_{ij}}^{z}=\frac{1}{4}\sum_{\sigma ,\nu }\eta \left(
\sigma \right) \eta \left( \nu \right) f_{\mathbf{r}_{ij},\sigma }^{\nu
\dagger }f_{\mathbf{r}_{ij},\sigma }^{\nu }$, where $\eta\left(\uparrow\right)=\eta\left(x\right)=+1$ and $\eta\left(\downarrow\right)=\eta\left(y\right)=-1$, and where the constraint of no double occupation has been used. 

Without loss of generality, we can assume that the ferromagnetic spin order
occurs along the $\hat{z}$ axis. Then $H_{H}^{d}$ can be written as

\begin{equation}
H_{H}^{d}=\frac{J}{2}\sum_{\mathbf{r}_{ij},\mathbf{a}}\left[ -S_{\mathbf{r}%
_{ij}}^{z}S_{\mathbf{r}_{ij}+\mathbf{a}}^{z}+3T_{\mathbf{r}_{ij}}^{z}T_{%
\mathbf{r}_{ij}+\mathbf{a}}^{z}+4U_{\mathbf{r}_{ij}}^{z}U_{\mathbf{r}_{ij}+%
\mathbf{a}}^{z}\right] +\rm{cst},
\end{equation}%
We now introduce the following Hubbard-Stratonovich (HS) decouplings

\begin{eqnarray}
-\frac{J}{2}\sum_{\mathbf{r}_{ij},\mathbf{a}}S_{\mathbf{r}_{ij}}^{z}.S_{%
\mathbf{r}_{ij}+\mathbf{a}}^{z}& \rightarrow \sum_{\mathbf{r}_{ij},\mathbf{a}%
}\left[ \frac{1}{2J}m_{\mathbf{r}_{ij}}m_{\mathbf{r}_{ij}+\mathbf{a}}+\frac{1%
}{2}m_{\mathbf{r}_{ij}+\mathbf{a}}S_{\mathbf{r}_{ij}}^{z}+\frac{1}{2}m_{%
\mathbf{r}_{ij}}S_{\mathbf{r}_{ij}+\mathbf{a}}^{z}\right] ,  \label{eq:HS_SS}
\\
\frac{3J}{2}\sum_{\mathbf{r}_{ij},\mathbf{a}}T_{\mathbf{r}_{ij}}^{z}T_{%
\mathbf{r}_{ij}+\mathbf{a}}^{z}& \rightarrow \sum_{\mathbf{r}_{ij},\mathbf{a}%
}\left[ -\frac{1}{6J}n_{\mathbf{r}_{ij}}n_{\mathbf{r}_{ij}+\mathbf{a}}+\frac{%
1}{2}n_{\mathbf{r}_{ij}+\mathbf{a}}T_{\mathbf{r}_{ij}}^{z}+\frac{1}{2}n_{%
\mathbf{r}_{ij}}T_{\mathbf{r}_{ij}+\mathbf{a}}^{z}\right] ,  \label{eq:HS_TT}
\\
2J\sum_{\mathbf{r}_{ij},\mathbf{a}}U_{\mathbf{r}_{ij}}^{z}U_{\mathbf{r}_{ij}+%
\mathbf{a}}^{z}& \rightarrow \sum_{\mathbf{r}_{ij},\mathbf{a}}\left[ -\frac{8%
}{J}p_{\mathbf{r}_{ij}}p_{\mathbf{r}_{ij}+\mathbf{a}}+4p_{\mathbf{r}_{ij}}U_{%
\mathbf{r}_{ij}+\mathbf{a}}^{z}+4p_{\mathbf{r}_{ij}+\mathbf{a}}U_{\mathbf{r}%
_{ij}}^{z}\right] ,  \label{eq:HS_UU}
\end{eqnarray}%
where $m_{\mathbf{r}_{ij}}$ is related to the usual magnetization order
parameter, $n_{\mathbf{r}_{ij}}$ is the orbital order parameter and $p_{%
\mathbf{r}_{ij}}$ is a magneto-orbital order parameter. In order for Eqs. (%
\ref{eq:HS_TT}) and (\ref{eq:HS_UU}) to be well-defined HS decouplings, the
fields $n_{\mathbf{r}_{ij}}$ and $p_{\mathbf{r}_{ij}}$ should have antiferromagnetic
order.  Introducing the Fourier decompositions

\begin{equation}
m_{\mathbf{r}_{ij}}=\sum_{\mathbf{q}}\frac{e^{i\left( \mathbf{Q}_{m}+\mathbf{q}\right) \mathbf{r}_{ij}}}{\sqrt{N}}m_{\mathbf{q}},\ \ n_{\mathbf{r}%
_{ij}}=\sum_{\mathbf{q}}\frac{e^{i\left( \mathbf{Q}_{n}+\mathbf{q}\right) \mathbf{r}_{ij}}}{\sqrt{N}}n_{\mathbf{q}},\ \ p_{\mathbf{r}_{ij}}=\sum_{\mathbf{q}}\frac{e^{i\left(\mathbf{Q}_{p}+\mathbf{q}\right) 
\mathbf{r}_{ij}}}{\sqrt{N}}p_{\mathbf{q}},  \label{fourier}
\end{equation}%
where $\mathbf{Q}_{m}=\left( 0,0\right) $ and $\mathbf{Q}_{n}=\mathbf{Q}_{p}=%
\mathbf{Q}=\left( \frac{\pi }{a_{0}},\frac{\pi }{a_{0}}\right) $, $H_{H}^{d}$
becomes  
\begin{eqnarray}
H_{H}^{d}&=&\sum_{\mathbf{q},\mathbf{a}}e^{-i\mathbf{q}\mathbf{a}}\left[ 
\frac{\beta }{2J}m_{\mathbf{q}}m_{-\mathbf{q}}-m_{\mathbf{q}}\sum_{\nu }%
\frac{1}{\sqrt{N}}\sum_{\mathbf{k},\omega _{n}}\frac{1}{2}\left( \bar{f}_{%
\mathbf{k},\uparrow }^{\nu }f_{\mathbf{k}-\mathbf{q},\uparrow }^{\nu }-\bar{f%
}_{\mathbf{k},\downarrow }^{\nu }f_{\mathbf{k}-\mathbf{q},\downarrow }^{\nu
}\right) _{\omega _{n}}\right]   \nonumber \\
&+&\sum_{\mathbf{q},\mathbf{a}}e^{-i\mathbf{q}\mathbf{a}}\left[ \frac{\beta 
}{6J}n_{\mathbf{q}}n_{-\mathbf{q}}-n_{\mathbf{q}}\sum_{\sigma }\frac{1}{%
\sqrt{N}}\sum_{\mathbf{k},\omega _{n}}\frac{1}{2}\left( \bar{f}_{\mathbf{k}%
,\sigma }^{x}f_{\mathbf{k}-\left( \mathbf{Q}+\mathbf{q}\right) ,\sigma }^{x}-%
\bar{f}_{\mathbf{k},\sigma }^{y}f_{\mathbf{k}-\left( \mathbf{Q}+\mathbf{q}%
\right) ,\sigma }^{y}\right) _{\omega _{n}}\right]   \nonumber \\
&+&\sum_{\mathbf{q},\mathbf{a}}e^{-i\mathbf{q}\mathbf{a}}\left[ \frac{\beta 8%
}{J}p_{\mathbf{q}}p_{-\mathbf{q}}-p_{\mathbf{q}}\sum_{\sigma \nu }\frac{1}{%
\sqrt{N}}\sum_{\mathbf{k},\omega _{n}}2\eta \left( \sigma \right) \eta
\left( \nu \right) \left( \bar{f}_{\mathbf{k},\sigma }^{\nu }f_{\mathbf{k}%
-\left( \mathbf{Q}+\mathbf{q}\right) ,\sigma }^{\nu }\right) _{\omega _{n}}%
\right] ,  \label{s1}
\end{eqnarray}%
The minimum of the free energy with respect to the order parameters  defines
the saddle-point equations

\begin{eqnarray}
m_{\mathbf{q}}&=&J\sum_{\nu }\frac{1}{\beta N}\sum_{\mathbf{k},\omega _{n}}%
\frac{1}{2}\left\langle \bar{f}_{\mathbf{k},\uparrow }^{\nu }\left( i\omega
_{n}\right) f_{\mathbf{k}+\mathbf{q},\uparrow }^{\nu }\left( i\omega
_{n}\right) -\bar{f}_{\mathbf{k},\downarrow }^{\nu }\left( i\omega
_{n}\right) f_{\mathbf{k}+\mathbf{q},\downarrow }^{\nu }\left( i\omega
_{n}\right) \right\rangle ,  \label{eq:mq_sp} \\
n_{\mathbf{q}}&=&3J\sum_{\sigma }\frac{1}{\beta N}\sum_{\mathbf{k},\omega
_{n}}\frac{1}{2}\left\langle \bar{f}_{\mathbf{k},\sigma }^{x}\left( i\omega
_{n}\right) f_{\mathbf{k}-\left( \mathbf{Q}-\mathbf{q}\right) ,\sigma
}^{x}\left( i\omega _{n}\right) -\bar{f}_{\mathbf{k},\sigma }^{y}\left(
i\omega _{n}\right) f_{\mathbf{k}-\left( \mathbf{Q}-\mathbf{q}\right)
,\sigma }^{y}\left( i\omega _{n}\right) \right\rangle ,  \label{eq:nq_sp} \\
p_{\mathbf{q}}&=&\frac{J}{4}\sum_{\sigma \nu }\frac{1}{\beta N}\sum_{\mathbf{%
k},\omega _{n}}2\eta \left( \sigma \right) \eta \left( \nu \right)
\left\langle \bar{f}_{\mathbf{k},\sigma }^{\nu }\left( i\omega _{n}\right)
f_{\mathbf{k}-\left( \mathbf{Q}-\mathbf{q}\right) ,\sigma }^{\nu }\left(
i\omega _{n}\right) \right\rangle .  \label{eq:pq_sp}
\end{eqnarray}%
In order to determine which of these order parameters is the first to
develop,
%(i.e., highest $T_{c}$) 
we compute the change of sign in the second
derivative of the free energy in the symmetric phase. This leads to the
following conditions for the critical value of $J$ for each instability
(similar to Stoner criteria)

\begin{equation}
\frac{1}{J_{c, m}}=\chi \left( \mathbf{0},0\right) ,\ \ \frac{1}{J_{c, n}}%
=3\chi \left( \mathbf{Q},0\right) ,\ \ \frac{1}{J_{c, p}}=4\chi \left( 
\mathbf{Q},0\right) ,\rm{ }  \label{jc}
\end{equation}%
where we have defined the static susceptibility of the electronic system,

\begin{equation}
\chi \left( \mathbf{q},0\right) \equiv \chi \left( \mathbf{q},\omega=0\right) =-\frac{4}{\beta N}\sum_{\mathbf{k},\omega
_{n}}\frac{G_{\nu ,\sigma }^{ff}\left( \mathbf{k},i\omega _{n}\right)
-G_{\nu ,\sigma }^{ff}\left( \mathbf{k}+\mathbf{q},i\omega _{n}\right) }{%
\epsilon _{\mathbf{k}}-\epsilon _{\mathbf{k}+\mathbf{q}}},
\label{eq:susceptibility}
\end{equation}%
and the pseudofermion Green function is given in \cite{mole} 
%by  
%\begin{equation}
%G_{\nu ,\sigma }^{ff}\left( \mathbf{k},i\omega _{n}\right) =\frac{1}{i\omega_{n}-E_{h}-\lambda -\epsilon _{\mathbf{k}}+z^{2}\frac{\Gamma }{\pi }\ln
%\left( \frac{i\omega _{n}-W}{i\omega _{n}+W}\right) },
%\end{equation}
%where $z=\left\langle b_{\mathbf{r}_{ij}}^{\dagger }\right\rangle $, $\lambda $ is a Lagrange multiplier used 
%to impose the constraint $z^{2}+\sum_{\sigma ,\nu =x,y}f_{\mathbf{r}_{ij},\sigma }^{\nu \dagger }
%f_{\mathbf{r}_{ij},\sigma }^{\nu }=1$, $W$ is half the band width and $\Gamma $ the resonant level width of the isolated impurity.   

\begin{figure}[t]
\begin{center}
\includegraphics[width=14pc]{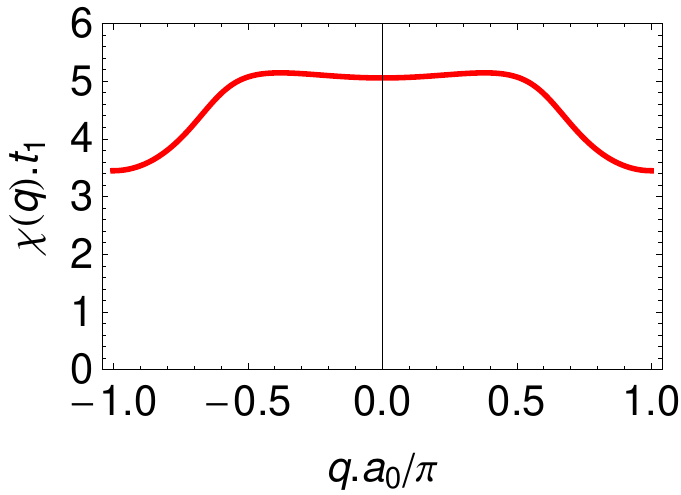}
%\hspace{2pc}
%\begin{minipage}[b]{14pc}
\caption{\label{chiq}Susceptibility $\protect\chi \left( \mathbf{q},0\right) $ vs $%
\mathbf{q}$ in the (1,1) direction. }
%\end{minipage}
\end{center}
\end{figure}

The problem now reduces to computing Eq. (\ref{eq:susceptibility}) for a
generic $\mathbf{q}$. Performing the Matsubara sum  at $T=0$ we obtain 
\begin{equation}
\chi \left( \mathbf{q},0\right) =\frac{4}{N}\sum_{\mathbf{k}}\frac{1}{\pi }%
\frac{\arctan \left( \frac{E_{h}+\lambda +\epsilon _{\mathbf{k}}}{%
z^{2}\Gamma }\right) -\arctan \left( \frac{E_{h}+\lambda +\epsilon _{\mathbf{%
k}+\mathbf{q}}}{z^{2}\Gamma }\right) }{\epsilon _{\mathbf{k}}-\epsilon _{%
\mathbf{k}+\mathbf{q}}},
\end{equation}
where $z=\left\langle b_{\mathbf{r}_{ij}}^{\dagger }\right\rangle $, $\lambda $ is a Lagrange multiplier used 
to impose the constraint $z^{2}+\sum_{\sigma ,\nu =x,y}f_{\mathbf{r}_{ij},\sigma }^{\nu \dagger }
f_{\mathbf{r}_{ij},\sigma }^{\nu }=1$, $W$ is half the band width and $\Gamma $ the resonant level 
width of the isolated impurity.   
This function is represented in Fig. \ref{chiq}

For $\mathbf{q}\rightarrow 0$ we recover the usual expression $\chi \left( 
\mathbf{0},0\right) =\rho _{f}^{t}\left( 0\right) $, where $\rho
_{f}^{t}\left( 0\right) $ is the total spectral density (summing over spin
and orbital) at the Fermi level. For the parameters $\lambda ,z$ that
minimize the SBMFA action, we obtain
$\chi \left( 0,0\right) =5.05/t_{1}$, $\chi \left( \mathbf{Q},0\right) =3.45/t_{1}$
which replaced in Eq. (\ref{jc}) lead to the following critical values ($t_{1}=0.007$ eV)

\begin{equation}
J_{c, m}\approx 16.1{\rm K}, \ \ J_{c, n}\approx 7.9{\rm K}, \ \ J_{c, p}\approx 5.9\ 
{\rm K}.  \label{eq:Jcm}
\end{equation}
We conclude that the first instability occurs in the magneto-orbital channel.

For FePc/Au(111) one can estimate $t_{1}=0.007$ eV, $%
t_{2}=3t_{1}$ and $U=1.6$ eV \cite{mole}. The exchange constant $J_{H}$ is
difficult to estimate for effective molecular orbitals. For pure Fe orbitals
it is of the order of 0.7 eV. From these estimates and the second Eq. (\ref%
{hd}), one expects $J \geq 1.8$ K. Thus, one might infer that the system
is not too far from an instability against spin ferro- and orbital
antiferro-magnetic order.

\section{RKKY interactions }

Another possible source of magnetic instabilities is the RKKY interaction,
which consists in the indirect interaction $I$ between spins mediated by
conduction electrons generated by the effective Kondo coupling between spins
and conduction electrons $J_{K}\mathbf{S}_{i}\cdot \mathbf{s}_{i}$ at second
order in $J_{K}$, where  $\mathbf{s}_{i}$ is the spin of the conduction
electrons at site $i.$

An advantage of Au and its (111) surface is that both the bulk sates and the
surface Shockley states near the Fermi energy can be described as free
electrons and therefore the calculations in Ref. \cite{kittel} for three
dimensions (3D) and Ref. \cite{beal} for the 2D case are valid. Following
these works one can write for dimension N=2 or 3 for two spins $\mathbf{S}%
_{1}$ and $\mathbf{S}_{2}$ at a distance $R$ (we will consider nearest
neighbors only)

\begin{equation}
H_{RKKY}\left( R\right) =I_{\rm{ND}}\left( R\right) \mathbf{S}_{1}.\mathbf{%
S}_{2}, \ \ \ \ I_{\rm{ND}}\left( R\right) =-\frac{1}{4}\tilde{J}_{\rm{ND}%
}^{2}\chi _{\rm{ND}}\left( R\right) ,  \label{hrkky}
\end{equation}%
where $\tilde{J}_{\rm{3D}}=J_{K}V_{\rm{at}}$ , $\tilde{J}_{\rm{2D}%
}=JS_{\rm{at}}$, $V_{\rm{at}}$ ($S_{\rm{at}}$) is the volume (surface)
per Au atom in the bulk (surface) and $\chi _{\rm{ND}}\left( R\right) $ is
the spin susceptibility, given by Eqs. (14) and (15) of Ref. \cite{beal}:
\begin{equation}
\chi _{\rm{3D}}\left( R\right) =-\rho _{\rm{3D}}\left( \epsilon
_{F}\right) \frac{4k_{F}^{3}}{\pi }F_{\rm{3D}}\left( 2k_{F}R\right), \chi _{\rm{2D}}\left( R\right) =-\rho _{\rm{2D}}\left( \epsilon
_{F}\right) k_{F}^{2}F_{\rm{2D}}\left( k_{F}R\right) ,  \label{sus}
\end{equation}%
where $\rho _{\rm{3D}}\left( \epsilon _{F}\right) =mk_{F}/2\pi ^{2}\hbar
^{2}$ ($\rho _{\rm{2D}}\left( \epsilon _{F}\right) =m^{\ast }/2\pi \hbar
^{2}$) is the density of states per spin and per unit volume (surface), and

\begin{equation}
F_{\rm 3D}\left( x\right) \equiv \frac{x\cos -\sin x}{x^{4}},\ \  F_{\rm 2D}\left( x\right) 
\equiv J_{0}\left( x\right) Y_{0}\left( x\right)
+J_{1}\left( x\right) Y_{1}\left( x\right) \xrightarrow[x \rightarrow\infty]{}
-\frac{\sin \left( 2x\right) }{\pi x^{2}},  \label{fn}
\end{equation}%
with $J_{\nu }\left( x\right) $($Y_{\nu }\left( x\right) $) the Bessel
function of the first (second) kind.

For the more realistic 3D case, using the value $k_{F}=1.21$ \AA$^{-1}$
for Au \cite{kevan}, one obtains $\rho _{\rm{3D}}\left( \epsilon
_{F}\right) =0.00805/$ (eV \AA$^{3}$). The density
per atom and spin projection is $\rho =\rho _{\rm{3D}}\left( \epsilon
_{F}\right) V_{\rm{at}}=0.137/$eV, where we have used $V_{\rm{at}}=17.0$ 
\AA$^{3}$ (the lattice parameter of f.c.c. Au is $a=4.08$ \AA  and 
$V_{\rm{at}}=a^{3}/4$). Keeping the product $\tilde{J}_{\rm{3D}}\rho _{\rm{%
3D}}\left( \epsilon _{F}\right) =J_{K}\rho $ that leads to the observed $%
T_{K}\simeq W\exp [-1/(2J_{K}\rho )]$ = 4.5 K, with $W=1/(2\rho )$ for the SU(4) impurity Kondo model, one obtains $%
J_{K}\simeq 0.4$ eV.

Using the above equations with $|F_{\rm{3D}}\left( x\right) |\leq 1/x^{3}$
for large $x$ and $R=14.7$ \AA for the
intermolecular distance \cite{tsuka} we obtain

\begin{equation}
|I_{\rm{3D}}|\leq \left( JV_{\rm{at}}\right) ^{2}\rho _{\rm{3D}}\left(
\epsilon _{F}\right) \frac{1}{\pi (2R)^{3}}=0.05\rm{ K.}  \label{i3}
\end{equation}

For 2D, the effective mass of the surface Shockley states is $m^{\ast }=0.28m
$ \cite{kevan}. This leads to $\rho _{\rm{2D}}\left( \epsilon _{F}\right)
=0.00589/\left( \rm{eV}\mathring{A}^{2}\right) $. Using $S_{\rm{at}}=%
\sqrt{3}a^{2}/4=7.21$ \AA$^{2}$ one obtains 
$\rho =\rho _{\rm{2D}}\left( \epsilon _{F}\right) S_{\rm{at}}=0.0425/\rm{eV}$. Knorr {\it et al.} 
have shown that the bulk states dominate the hybridization with the
impurity
%and therefore are more important than the surface states in the Kondo screening 
\cite{knorr}. Assuming (as an overestimation) that half of
the contribution to $J_{K}\rho $ is due to surface states 
%(and the other half to bulk states) 
leads to $J_{K}\simeq 0.65$ eV. From Eq. (\ref{fn}) for
large $x$, $|F_{\rm{2D}}\left( x\right) |\leq 1/\left( \pi x^{2}\right) $,
and using Eqs. (\ref{hrkky}) and (\ref{sus}) for $R=14.7$ \AA  we obtain

\begin{equation}
|I_{\rm{2D}}|\leq \left( JS_{\rm{at}}\right) ^{2}\rho _{\rm{2D}}\left(
\epsilon _{F}\right) \frac{1}{4\pi R^{2}}=0.54\rm{ K.}  \label{i2}
\end{equation}

A calculation that follows the same steps as done in the previous section
shows that to have a ferromagnetic instability in the system $I_{\rm{ND}}<0
$, with $|I_{\rm{ND}}|>I_{c}=J_{cm}\approx 16.1$ K. Therefore, magnetic
instabilities driven by the RKKY interaction are unlikely for FePc/Au(111). 

\section{Summary and discussion}

We have studied the magnetic and orbital instabilities of a model used before to explain
the scanning tunneling spectroscopy (STS) of a system of FePc molecules on Au(111).
The model generalizes to the lattice the SU(4) Anderson model and is expected to have emergent
SU(4) symmetry at low energies.

We find that due to effective generalized exchange interactions originated by the hopping terms, 
the system is close to a combined instability of spin ferromagnetic and orbital antiferromagnetic 
character. Due to this combined character it is possible that the application of a
magnetic field induces not only a finite magnetization but also a checkerboard orbital
ordering, which might be observed by STS if the tip is not radially symmetric.

\section*{Acknowledgments}

AML acknowledges support form JQI-NSF-PFC. AAA is partially supported by CONICET, Argentina. 
This work was sponsored by PICT 2010-1060 and 2013-1045 of the ANPCyT-Argentina and 
PIP 112-201101-00832 of CONICET.

\end{document}